# Efficient Modelling of Ion Structure and Dynamics in Inorganic Metal Halide Perovskites

Salvador Rodríguez-Gómez Balestra†[a], Jose Manuel Vicent-Luna†[b], Sofia Calero[a], Shuxia Tao[b,*], Juan A. Anta[a,*]

Metal halide perovskites (MHPs) are nowadays one of the most studied semiconductors due to their exceptional performance as active layers in solar cells. Although MHPs are excellent solid-state semiconductors, they are also ionic compounds, where ion migration plays a decisive role in their formation, their photovoltaic performance and their long-term stability. Given the above-mentioned complexity, molecular dynamics simulations based on classical force fields are especially suited to study MHP properties, such as lattice dynamics and ion migration. In particular, the possibility to model mixed compositions is important since they are the most relevant to optimize the optical band gap and the stability. With this intention, we employ DFT calculations and a genetic algorithm to develop a fully transferable classical force field valid for the benchmark inorganic perovskite compositional set $CsPb(Br_xI_{1-x})_3$ ($x$ = 0, 1/3, 2/3, 1). The resulting force field reproduces correctly, with a common set of parameters valid for all compositions, the experimental lattice parameter as a function of bromide/iodide ratio, the ion-ion distances and the XRD spectra of the pure and mixed structures. The simulated thermal conductivities and ion migration activation energies of the pure compounds are also in good agreement with experimental trends. Our molecular dynamics simulations make it possible to predict the compositional dependence of the ionic diffusion coefficient on bromide/iodide ratio and vacancy concentration. For vacancy concentrations of around $9 \times 10^{21}$ cm$^{-3}$, we obtained ionic diffusion coefficients at ambient temperature of $10^{-11}$ and $10^{-13}$ cm$^2$/s for $CsPbBr_3$ and $CsPbI_3$, respectively. Interestingly, in comparison with the pure compounds, we predict a significantly lower activation energy for vacancy migration and faster diffusion for the mixed perovskites. This anomalous effect could be of relevance for the design of perovskite alloys with optimized performance and requires further exploration.

## Introduction

Perovskite solar cells (PSCs) are an emerging photovoltaic technology that has attracted huge interest from researchers, technologists and investors. PSCs have an impressive learning curve, which has reached a certified record efficiency of 25.2%[1] in less than a decade.[2] The active light absorber material in PSCs are metal halide perovskites (MHPs), which are mixed ionic-electronic semiconductors with exceptional optoelectronic properties.[3–6] MHPs have the generic $ABX_3$ chemical formula, where A is a relatively large organic or inorganic monovalent cation (typically for solar cells A is $Cs^+$, $CH_3NH_3^+$, or $(CH(NH_2))^+$), B is a metal divalent cation (typically B = $Pb^{2+}$) and X a halide anion (X = $Br^-$ or $I^-$, and sometimes $Cl^-$). To improve the performance and stability of the materials and the devices, state-of-the-art PSCs with record efficiencies[7,8] and improved lifetimes[9] were mostly accomplished by using mixed compositions of the A and X ions. For instance, a typical strategy is to mix Br and I in MHPs to tune the band gap and maximize simultaneously the open circuit photovoltage and the short-circuit photocurrent of the final device.[10]

In spite of the huge potential of MHPs, they are very complex materials. Due to their mixed ionic-electronic character, ion migration,[11] phase segregation,[12] and degradation phenomena[13,14] are commonplace in actual PSC tests and compromise critically the reproducibility and the performance of future developments. The main problem of studying these processes and their impact on photovoltaic performance is that they occur in relatively long-time scales. Molecular simulation and modeling are very useful tools to understand the properties and behavior of functional materials from a fundamental perspective. Density Functional Theory (DFT) approaches have been extensively used to describe MHP and provided very useful insights.[15,16] However, DFT is computationally expensive and its sphere of applicability is constrained to small samples (a few unit cells) and very short time scales (a few ps).[17] This makes it difficult to study MHP alloys, which require larger samples, and ionic migration, for which longer simulations are needed. To estimate for instance an ionic diffusion coefficient from DFT and indirect approach should be followed, first calculating the diffusional energy barrier for ions and then applying the Arrhenius formalism.[11,18]

As an alternative to *ab initio* methods such as DFT, Classical Molecular Dynamics (CMD) is computationally cheaper and offers the possibility to simulate larger samples and to cover longer simulation times. The critical drawback of CMD studies, as regards their predictive power, is that one needs a suitable and realistic force field. Classical force fields have already been developed to model MHPs and to predict successfully ionic diffusion coefficients and phase transition temperatures.[19,20] These force fields include Lennard-Jones plus coulombic terms. Force fields of this type have been found to correctly reproduce the first stages of perovskite crystal formation and dynamics in the liquid phase.[21]

For solid-state perovskites Mattoni and coworkers[19,22] developed a force field for hybrid MHP ($CH_3NH_3PbI_3$) that combined Buckingham, Lennard-Jones and coulombic terms, for the all-inorganic interactions and the usual functional forms for bonds, bends, and torsions of the intramolecular interactions used in the $CH_3NH_3$ molecule. This force field is

[a.] Área de Química Física, Universidad Pablo de Olavide, 41013 Seville, Spain, e-mail: anta@upo.es
[b.] Center for Computational Energy Research, Department of Applied Physics, Eindhoven University of Technology, P.O. Box 513, 5600MB, Eindhoven, The Netherlands; e-mail: S.X.Tao@differ.nl
† These two authors contributed equally to this work.

suitable to model solid MHPs because it combines simplicity and accuracy in a very convenient manner. Thus, large samples of up to 4 unit cells (>3000 atoms) and long simulation times of up to a ten nanoseconds can be studied.[19] Despite the active development in new force fields for MHPs (as mentioned above), the available ones are mostly tailored to simulate pure compounds. Universal force fields capable of simulating mixed compounds are rarely seen.

A classical force field contains fitting parameters. A convenient choice to find a good estimation for them is to use quantum DFT calculations structures and energies as a reference to tune the classical potential parameters so that quantum observables can be reproduced. This was the strategy followed by Mattoni et al.[19] Thus, they used hydrostatic deformations of the unit cell and the dependence of the lattice energy on the orientation of the $CH_3NH_3^+$ organic cation to fit the classical force field parameters. This way phase transition temperatures, ion migration activation energies and rotational relaxation times of $CH_3NH_3PbI_3$[19] and $CH_3NH_3PbBr_3$[20] were reasonably well reproduced. The same strategy was used by Qian et al.[23] for the same functional form and the resulting potential predicted quite well the thermal conductivity and the cubic-tetragonal transition of $CH_3NH_3PbI_3$.[24] Handley and Freeman[25] developed a similar force field but with parameters that are shared with the perovskite precursors methylammonium iodide and lead iodide. Their force field reproduces correctly the phonon density of states and cation dynamics. Jinnouchi and coworkers[26] used instead Machine Learning to fit the potential parameters "on the fly" (at the same time that the simulation runs). The resulting method reproduced adequately the experimental phase transitions of the $CH_3NH_3PbI_3$ perovskite. The formation and migration of iodine vacancies in $CH_3NH_3PbI_3$ have equally been studied by means of the MYP force field by Barboni and Souza.[27] These authors exploited the advantages of CMD to describe how vacancy concentration and microscopic structure impacts ionic conduction in MHPs.

There is a main issue with the classical force fields developed so far for MHPs, which is the lack of transferability and universality of the parameters. This is for instance evidenced by the fact that the fitted parameter for Pb is not the same in $CH_3NH_3PbI_3$[19] as in $CH_3NH_3PbBr_3$.[20] However, there are indications that electron density of Pb is essentially the same in both perovskites.[28] The lack of transferability is also evidenced when the fitted parameters reproduce only a very small set of configurations or material properties, like for instance the isotropic deformation of a cubic cell.[20] These issues complicate the ability of the CMD modeling to reproduce adequately the static and the dynamic properties of either mixed compositions or in a wide range of experimental conditions, which is what is precisely needed for photovoltaic optimization.

In this work we make use of a genetic algorithm to fit the force field parameters to DFT energies calculated for a very wide set of configurations. In contrast to previous developments, we use a large set of deformations (not only isotropic) to perform the fitting. To test our method we choose the all-inorganic perovskite $CsPb(Br_xI_{1-x})_3$, with $x$ ranging between 0 and 1. This covers the range of compositions commonly used in the experiments to optimized performance. Our focus is to devise a force field that combines low computational cost (needed to run large-sample long-scale simulations) with versatility and transferability for mixed perovskites. We have tested this by looking at the ability of the force field to reproduce the experimental structural parameters, that is, lattice constants, mechanical and thermal properties, XRD spectra and ion migration activation energies, of a range of mixed compositions with a single parameter set. We also report interesting non-linear dynamical properties of the mixed perovskites. Our work demonstrates the possibility of developing transferable force fields for large scale MD simulations and its potential in discovering and designing new dynamic properties of MHPs.

## Methodology

### Step 1. Force field development

Due to the inorganic nature of the $CsPb(Br_xI_{1-x})_3$ perovskites, we use in this work a functional form including the Buckingham and Coulomb terms:

$$U_{ij}(r_{ij}) = A_{ij} \exp\left(-\frac{r_{ij}}{\rho_{ij}}\right) - \frac{C_{ij}}{r_{ij}^6} + \frac{1}{4\pi\varepsilon_0}\frac{q_i q_j}{r_{ij}} \quad (1)$$

where $r_{ij} = \|\vec{r_{ij}}\|$ is the distance between the $i$ and $j$ atoms, $q_i, q_j$ are their respective atomic point charges and $A_{ij}, \rho_{ij}, C_{ij}$ are the three parameters of the Buckingham potential.

The reference data used for the fitting of the force field parameters were obtained from Density Functional Theory (DFT) calculations on the periodic structures of the pure and mixed perovskites. For that, a set of geometries were generated by producing small deformations of the perovskite cell around the minimum energy structure. The generated geometries and corresponding DFT energies are the observables used to minimize the cost function[29–31]

$$F = \sum_i w_i (E_i - E_i^{obs})^2 \quad (2)$$

where $E_i$ is the energy of the geometry as produced by the classical force field, $E_i^{obs}$ is the corresponding "observable" obtained from the DFT calculations for this geometry, and $w_i$ is the weighting factor that this structure has in the whole set of selected structures. We use the Boltzmann factors of the energy difference of each structure with respect to the minimum energy configuration for each composition as the weighting factors in Eq. (2), that is, $w_i = \exp(-(E_i - E_{min})/k_B T)$.[32] Because of the Boltzmann distribution, only geometries with small distortions around the minimum energy configuration will be significant in the generation of force field parameters. We choose ambient temperature 300 K as the temperature in the Boltzmann factor. Temperatures higher than this favor the stabilization of the cubic structure, whereas at lower temperatures the orthorhombic phase tends to be dominant, as reported in the literature.[33–35]

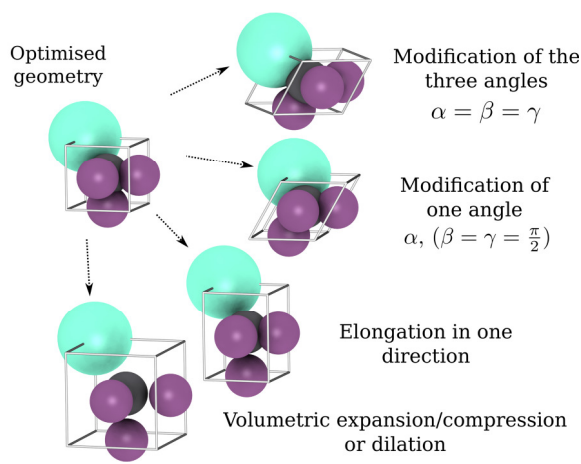

**Figure 1.** Distortions implemented in perovskite structures for energy calculations (Eq. (2))

The geometries used to generate the reference data with which to optimize the $F$ function were generated with an in-home code. The code produces a distortion in the cell in a chosen direction (*a*, *b*, *c*). The applied distortions are illustrated in **Figure 1**: linear expansion (the volume is not fixed), shear deformations (deformation in angles but keeping the cell volume fixed), distortions applied in one axis only (keeping the total cell volume fixed) and different distortions applied in the three axes. By applying all distortions, we have generated over 1000 geometries. The DFT calculations were performed using the Vienna *Ab initio* Simulation Package (VASP),[36] employing pseudo-potentials within the projector-augmented (PAW) wave method,[37] and the Perdew, Burke and Ernzerhof (PBE) exchange-correlation functional.[38] Calculations were run with a cut-off energy of 500 eV and convergence thresholds of 0.1 meV and 20 meV Å$^{-1}$ for energy and force, respectively. Ions were allowed to move in this structural optimization stage of the perovskite simulation cell.

Dispersion energy contributions were treated with the methods of Grimme (D3-Grimme),[39] and Tkatchenko-Scheffler.[40] The atomic charges used in the interatomic potential have been derived from the DFT calculations by the Hirshfeld-I procedure, since this is the default procedure in the Tkatchenko-Scheffler method which we have used to calculate the reference DFT energies. It is important to note that they vary slightly with composition and geometry.[41] However, to accomplish a universal and transferable force field, the calculated atomic charges were averaged over the values derived from each composition and geometry so that a common set is obtained. Alternatively, we have used other two methods to derive the point charges: both methods predict the same charge relation among atomic types. However, the Bader's method[42] yields a scaling factor of 0.82-0.80 (for Pb and I atoms) respect the Hirshfeld-I procedure, whereas the DDEC6 method[43,44] yields 0.92-0.86.

In this work we have developed a genetic algorithm to fit the force field parameters to the DFT energies of the configurations generated as described above. We introduce the so-called FLAMA code (**F**orcefie**L**d **A**djust**M**ent **A**lgorithm), which is a modification of the GAIAST code[45] originally developed to obtain multicomponent isotherms within the framework of the Ideal Adsorption Solution Theory (IAST).[46] In general, a genetic algorithm provides the *genome*, that is, the set of parameters that minimizes the cost function, by mimicking natural selection and survival of the fittest. GAIAST and FLAMA algorithms are designed to optimize continuous and non-ranged variables taking into account the physical constraints of the parameters. FLAMA makes internal calls to the GULP simulation package[47] to calculate the energy of each geometry using the force field from Eq. (1) with a test set of parameters. The IEEE 754 Standard is implemented by encoding the set of interatomic parameters into a one binary 32 number (called "phenotype" and "genotype", respectively). Three standardized genetic operators (*crossover*, *mutations*, and *elitism*), are used to evolve the phenotype and optimize the cost function of Eq. (2) until convergence is achieved.

The phenotype is initially composed of thirty force field parameters corresponding to the three parameters $A_{ij}$, $\rho_{ij}$, $C_{ij}$ in Eq. (1) for *i-j* atom-atom combinations. Point charges are excluded from the genetic procedure as their values are taken from the quantum procedure as indicated above (and then averaged over the various configurations and compositions). Note also that some parameters were taken zero since their contribution is negligible (this is for instance the case of the Pb-Pb interaction, which is predominantly coulombic). Importantly, the determination of the final phenotype is based on the restriction that the force field should be transferable/universal, *i.e.*, that it is ideally valid for all compositions and geometries of CsPb(Br$_x$I$_{1-x}$)$_3$, with *x* ranging between 0 and 1. With this intention, all compositions are considered in the same pool of energies, so the genetic algorithm seeks a common phenotype. However, the procedure is started by optimizing the phenotype of the pure perovskites and then extended to the mixed configurations following an iterative procedure. We have obtained better fits for pure structures than for the mixtures. This is due to the method used and to the effect of the strong coupling between the $A_{ij}$ and $\rho_{ij}$ parameters of the Buckingham potential (the optimization process tends to get stuck with multiple local minimums far from the global minimum). Nevertheless, fitted parameters have been accomplished for all considered geometries. As will be discussed later, the obtained force field describe with a fair degree of accuracy all compositions. A flowchart with the basic structure of the FLAMA algorithm is presented and described in **Scheme 1**. The result of this genetic optimization is presented in **Table 1**. The code FLAMA is open access in the public repository: http://github/salrodgom/flama.

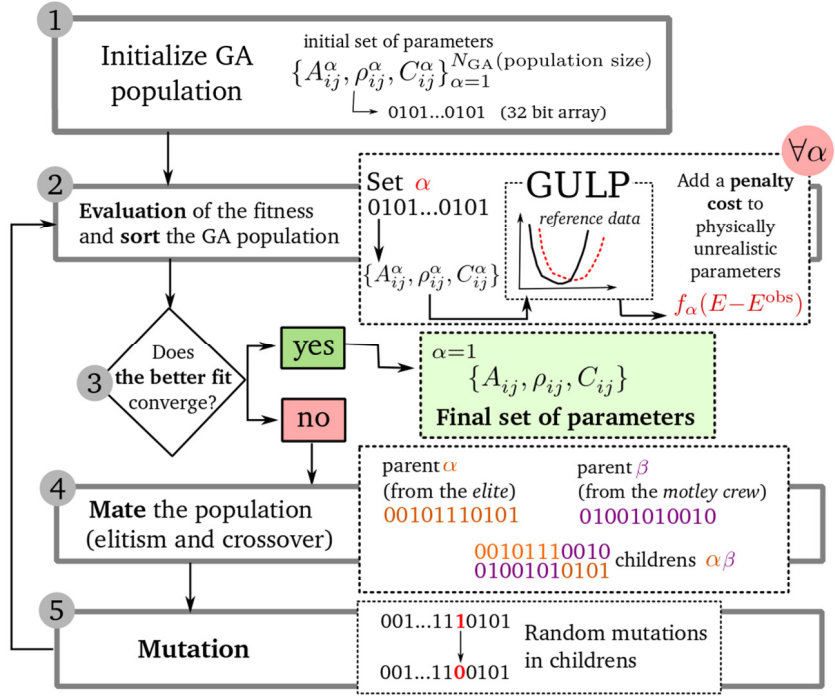

**Scheme 1.** FLAMA algorithm. **Box 1:** initial population of genome ($N_{GA}$ = 100). The first 15% of the initial population is composed of reference literature data (adaptations of generic force fields, theoretical values of ionic and vdw radius, $C_6$ values from DFT, etc.) The other 85% of the initial values are randomly generated. Then the entire population is converted in a 32×3×$N_{GA}$ array of 0's and 1's. **Box 2:** Evaluation and sorting in *classes* of the cost function of each parameter set according to Eq. (2) for the energy-geometry pairs indicated in Figure 1. Force field-based energies are calculated using an internal interface with the GULP code. Reference data are taken from DFT. A penalty cost is added for all configurations with physically unrealistic parameters. Classes: 15% of the population with the lowest cost function is labeled "best". 5% best is the "elite". The 85% worst is the "motley crew". **Box 3:** Convergence test of the parameter set with lowest cost function (according to a tolerance factor). If the test is kept within 15 iterations the algorithm is stopped and the final parameter set is obtained. **Box 4**: Mating of population with the *crossover* and *elitism* operators. Crossovers mate randomly genomes of different *classes:* for the 15% best of the population (except 5% "elite"), parents are replaced by offspring which are mixtures of the "best" and the "elite". The "motley crew" is mixed with the "best" to bring it to date. Elistism duplicates the population of the elite for the next generation. The total population $N_{GA}$ is preserved for all steps. With this scheme the "elite" class maintains stability while the "motley crew" introduces the necessary changes to evolve to the best configuration (lowest cost function). **Box 5:** random mutations introduced on the children. Return to **Box 2**.

**Table 1.** Fractional point charges considered in the genetic algorithm and resulting force field potential parameters for CsPb(Br$_x$I$_{1-x}$)$_3$ perovskites

| Point charges | | Buckingham interactions | | | |
| --- | --- | --- | --- | --- | --- |
| Species | $q_i$ /e | Pairs | $A_{ij}$ / eV | $\rho_{ij}$ / Å | $C_{ij}$ / eV Å$^6$ |
| Pb | +0.9199 | Pb – Pb | 0 | 0.2500 | 1024 |
| Cs | +1.0520 | Pb – Cs | 8753 | 0.4187 | 1.108 |
| I | -0.6573 | Pb – I | 8955 | 0.3040 | 267.5 |
| Br | -0.6573 | Pb – Br | 6447 | 0.3010 | 330.9 |
| | | Cs – Cs | 7512 | 0.04378 | 227.8 |
| | | Cs – I | 40322 | 0.3310 | 0 |
| | | Cs – Br | 9345 | 0.3331 | 117.5 |
| | | I – I | 62090 | 0.2752 | 1022 |
| | | I – Br | 151700 | 0.2663 | 918.4 |
| | | Br – Br | 370700 | 0.2420 | 277.9 |

**Step 2. Force-field based Molecular Dynamics simulations**

To test the accuracy and reliability of the force field parameters determined in Step 1, we obtain structural and dynamic properties using CMD simulations. Prior to this it is necessary to find the equilibrium density and geometry at a given temperature. To do that we carried out simulations in the NPTPR ensemble (also known as NσT ensemble) between 300 K and 700 K at 1 bar. In these simulations the three cell parameters are allowed to vary independently while the three angles remain fixed at right angles so the equilibrium density at a given pressure is reached. These simulations consisted of $2 \times 10^5$ steps with a time step of 0.5 fs, ensuring that the volume and the energy of the systems fluctuates over time around an average value at each temperature. The simulation box contains 6×6×6 and 4×3×4 unit cells for the cubic and the orthorhombic perovskites, respectively. The pressure and temperature were controlled by the Martyna-Tuckerman-Tobias-Klein barostat[48] (dumping parameter 100 steps) and the Nosé-Hoover thermostat (dumping parameter 1000 steps),[49] respectively.

For simplicity, calculation of transport properties of the studied perovskites was restricted to the cubic structures. Once all the structures have been relaxed to their equilibrium volume at each temperature, 12 or 24 iodide or bromide atoms were removed to create vacancies. The concentration of vacancies depends on the equilibrium volume of the system, which is different for each chemical composition and temperature. 12 and 24 vacancies correspond to approximate vacancy concentrations of $(8.9 \pm 1.2) \times 10^{21}$ and $(17.8 \pm 2.4) \times 10^{21}$ cm$^{-3}$, respectively. We checked that the formation of the vacancies does not affect to the equilibrium volume of the systems by performing an additional NPTPR simulation. To compensate for the charge of the missing halide atoms the charge of the lead atoms was reduced by a uniform charge as described in the literature.[27,50] Once temperature and density are determined by the NPTPR simulations, production runs in the NVE ensemble with a total simulation time of 10 ns ($2 \times 10^7$ steps) were carried out. CMD simulations were performed with the RASPA (NVE simulations)[51] and LAMMPS molecular simulation software.[52] The thermal conductivity was calculated using the OCTP plug-in for LAMMPS.[53]

Radial distribution functions were computed from thermal averaging of the atomic positions in the production run.[54] Ionic diffusion coefficients were obtained via the computation of the Mean Square Displacement (MSD) of particles, defined as a statistical average of their relative positions in the simulation

$$\langle (r(t) - r(0))^2 \rangle = \frac{1}{N}\sum_{i=1}^{N} \|\vec{r}(t) - \vec{r}(0)\|^2 \quad (3)$$

where $i$ is an index that runs over the total number of particles of each type $N$. In this work we monitor the displacement of all halide ions. This choice is based on the assumption that ionic conduction in perovskites is mainly originated from the migration of halide vacancies, at least at short times (up to seconds).[11,18] The MSD has the following time dependence for three-dimensional system[55]

$$\langle (r(t) - r(0))^2 \rangle = 6D_s t^\alpha \quad (4)$$

where $\alpha$ is a power exponent that reflects the nature of the atomic motion: for normal diffusion $\alpha = 1$ whereas $\alpha \neq 1$ is an indicative of anomalous diffusion.

**Results and discussion**

**A: Lattice energies and force field generation**

Step 1 in the methodology used a genetic algorithm to find a set of operational parameters for perovskites of generic formula CsPb(Br$_x$I$_{1-x}$)$_3$ based on the classical force field (Eq. (1)) **Table 1** collects the results of this procedure for the point charges and the interatomic parameters $A_{ij}$, $\rho_{ij}$, $C_{ij}$. It can be seen that point charges are closer to the nominal oxidation states of the constituent ions that in previous force field sets.[19] The departure from the nominal oxidation state is a measure of the covalent character of the bond. Not surprisingly, this is largest for Pb$^{2+}$ and smallest for Cs$^+$. Importantly, the force field parameters are equal for all values of the stoichiometric parameter $x$. This desirable property, i.e., transferability, appears to be supported by previous quantum calculations,[28] at least for the exchange of Br and I.

**Table 2.** Comparison between the calculated elastic constants with the force field and reference data. Calculations in this work are done by structural optimization (0 K) and using the Voight's convention with the GULP code.

| Composition | Bulk moduli / GPa | Young's Modulus / Ga | Shear modulus / GPa | Poisson's ratio | References |
|---|---|---|---|---|---|
| $CsPbI_3$ (room temperature and non-perovskite symmetry) | 19.8 | 20.1 | 7.9 | 0.33 | Ref. [56] |
| $CsPbI_3$ | 14.38 | 14.49 | 6.3 | 0.32 | Ref. [57] |
|  | 16.55 | 30.78 | 9.93 | 0.19 | This work |
|  | 15.5 | 15.8 | 5.9 | 0.33 | Ref. [56] |
| $CsPbBr_3$ | 18.45 | 20.74 | 7.9 | 0.31 | Ref. [57] |
|  | 23.5 |  |  |  | Ref. [28] |
|  | 21.19 | 28.38 | 12.71 | 0.27 | This work |
| $CsPbBr_2I$ | 19.3 | 26.0 | 11.58 | - | This work |

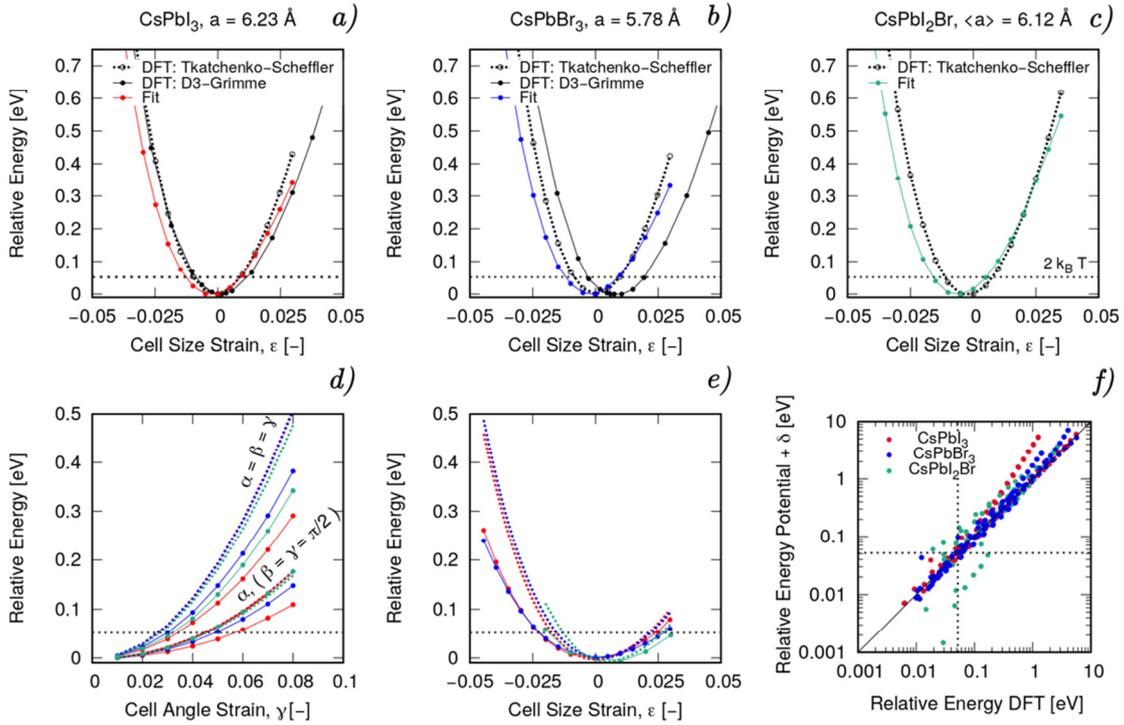

**Figure 2.** Relative energies of $CsPb(Br_xI_{1-x})_3$ simulated using the genetically-optimized force field compared with those from DFT. From left to right, from top to bottom in this order: (a) volumetric expansion (constant cell parameters α β and γ) with Cauchy strain $\varepsilon = \Delta a/a$ for $CsPbI_3$ composition (solid red line the force field, solid black line DFT calculation using the D3-Grimme method, and dashed black line the DFT reference data using the Tkatchenko-Scheffler method) (b) $CsPbBr_3$ composition (c) $CsPbI_2Br$ composition, (d) non-pure shear distortions (by changing $\alpha = \beta = \gamma$, and α by keeping fixed $\beta = \gamma = \frac{\pi}{2}$) with strain $\gamma = 1 - 2\alpha/\pi$, (the code color is the same than in the Top Figures) (e) anisotropic deformation in x- direction only, and (f) Shifted lattice energies of the compositions of perovskite using the genetically-optimized interatomic potential of this work vs. the DFT reference data. The horizontal dashed black lines (and vertical in f-subfigure) marks values at 600 K in all plots.

Figure 2 compares the lattice energies from the genetically-optimized classical force field and the DFT calculations for three types of geometrical distortions. It is worth mentioning the difference between the predictions of the D3-Grimme and Tkatchenko-Scheffler methods. Both methods give essentially the same results, however for the $CsPbBr_3$ composition the D3-Grimme method predicts a slightly larger cell parameter than that of the Tkatchenko-Scheffler method. The force field with genetically-optimized parameters accurately describes the energy penalty of deforming isotropically the cell (volumetric expansion or dilatation). Larger deviations are found for anisotropic distortions (shear and one-axis only elongation) although the minimum energy configuration is accurately located by the genetically-optimized force field in all cases. The calculation of the energy penalty is consistent with the exact description of the elastic and mechanical constants, as well as thermal conductivity, as compared with literature values (**Table 2**). Elastic and mechanical constants at the zero-point state were computed with the GULP code.

The relative deviations shown in Figure 2 demonstrate that the genetically-optimized classical force field reproduces remarkably well the DFT energies for a very wide range of applied strains of the crystal lattice, both isotropic and anisotropic. Relatively large errors are however found for the mixed perovskites, particularly in the low deformation zone. This is expected as the genetic algorithm was initially applied to the pure compositions and then iteratively extended to the mixtures. Bear in mind that for simplicity and numerical efficiency not all possible configurations of the mixed perovskites were explored. Hence, there is room for improvement in the optimization of the force field, although the remarkable agreement with experimental data found in the next sections suggest that the figures presented in Table 1 can describe efficiently these perovskites.

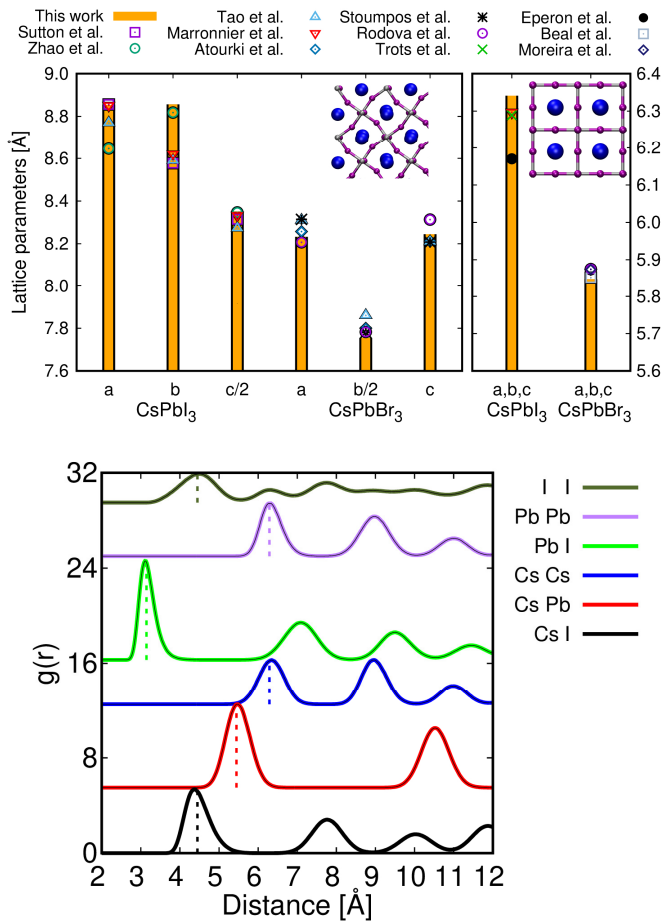

**Figure 3.** Structural parameters at ambient conditions (300 K and 1 bar) **Top:** Lattice parameters of the pure perovskites $CsPbBr_3$ and $CsPbI_3$ obtained in this work and taken from several literature reports.[33–35,58–65] **Bottom:** CMD computed radial distribution functions of the $CsPbI_3$ perovskite. Dashed lines stand for experimental interatomic distances in the cubic form.[58]

### B: Structural parameters of pure perovskites

Once developed the optimized force field, the next step it to test its ability to reproduce structural and dynamical properties of the full compositional set $CsPb(Br_xI_{1-x})_3$, with $x$ ranging between 0 and 1. Structural parameters for pure perovksites are collected in **Figure 3**. The lattice parameters of $CsPbBr_3$ and $CsPbI_3$ as predicted by the CMD simulation at ambient conditions (300 K and 1 bar) compare remarkably well with data taken from the literature. Bearing in mind the high dispersion of experimental results (especially for the orthorhombic phase of $CsPbI_3$), the genetically-optimized classical force field is accurate enough to reproduce the reported crystallographic data. This accuracy is confirmed when the computed radial distributions functions are compared with experimental[58] interatomic distances.

The genetically-optimized classical force field also yields reasonably accuracy in the thermal expansion of the lattice compared with experiments for $CsPbBr_3$ and $CsPbI_3$ (**Figure 4**). Although the simulation does not show a transition to a more stable orthorhombic structure at low temperatures (as observed in the experiments), the thermal expansion of the normalized lattice parameter in the full temperature range of 400 K is remarkably well reproduced for both the iodide and the bromide perovskite. It should be noted that the experiments at low temperatures provide conflicting results, especially for $CsPbI_3$.[58,62] This may have to do with the fact that the orthorhombic phase is metastable and can only be (pseudo)stabilized when the system is cooled down very slowly.[62] It is therefore not surprising that the classical force field, genetically optimized for the cubic structure, is not capable of predicting this (pseudo)transition.

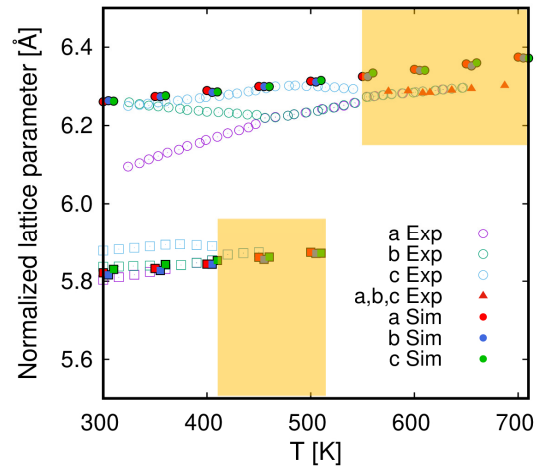

**Figure 4.** Lattice parameter as a function of temperature for $CsPbBr_3$ (bottom) and $CsPbI_3$ (top). Experimental data are taken from Ref.[62] The yellow-shadowed areas correspond to the regions with stabilized crystal structures according to the experiments. Note that the lattice parameters of the orthorhombic structure have been normalized as $a \rightarrow a/2^{1/2}$, $b \rightarrow b/2^{1/2}$ and $c \rightarrow c/2$

### C: Structural parameters of mixed perovskites

One of the most necessary features of any force field is the ability to describe the properties of a class of materials for a variety of compositions and different experimental situations, i.e. it must be transferable. We next investigate the ability of the our force field in predicting the whole range of the compositions for $CsPb(Br_xI_{1-x})_3$. **Figure 5** compares the simulated and experimental lattice parameters and the XRD spectra of the cubic and orthorhombic structures as a function of bromine content.[60,64] Results for both cubic and orthorhombic structures are included in the graph.

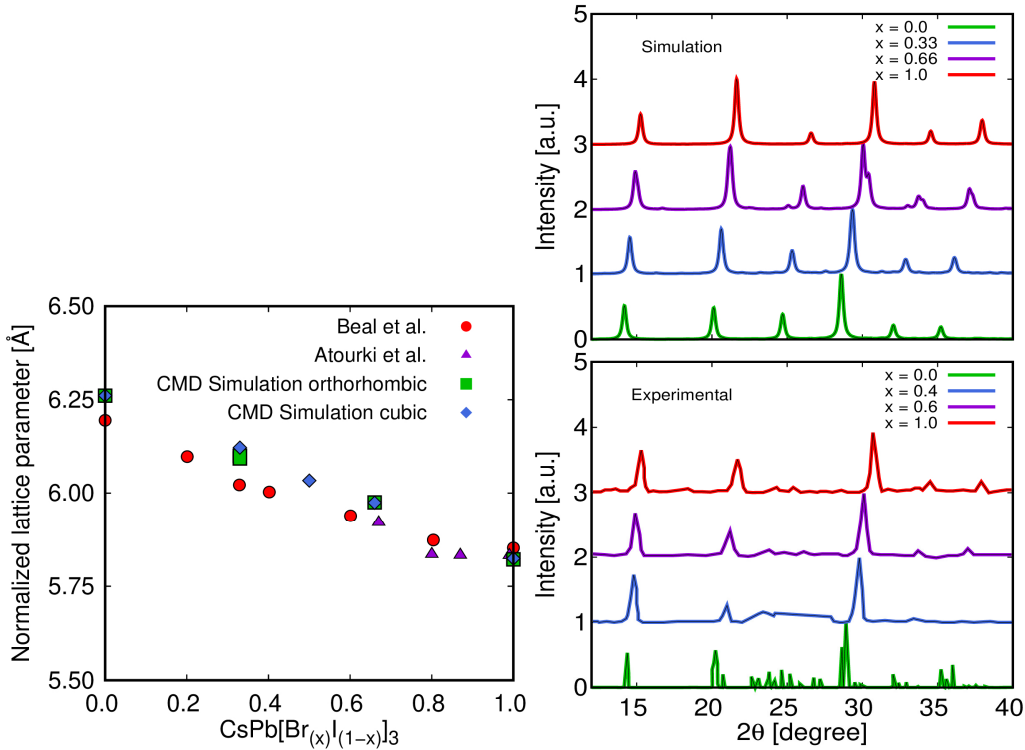

Figure 5. Structural parameters predicted by the genetically-optimized classical force field as a function of compositional parameter x and experimental data. **Left:** Lattice parameters. Experimental data are taken from Refs.[60,64] Note that the lattice parameter a of the orthorhombic structure has been normalized by a → a/2$^{1/2}$ **Right:** Simulated (top) and experimental (bottom) XRD spectra for CsPb(Br$_x$I$_{1-x}$)$_3$ and several values of the compositional parameter x. Experimental data taken from Refs.[60,61]

The most important result is that the classical potential with genetically-optimized parameters describes with remarkable precision the lattice compression upon iodine substitution by bromine. The predicted variation is essentially identical for the cubic and the orthorhombic structures, and both are very close to the experiments. Transferability of the force field over compositional variation is confirmed when theoretical XRD spectra from the CMD simulation are compared with measurements. The simulated spectra catch with precision the position of the main reflections of the lattice at all compositions. Hence, the classical force field reproduces quite well the expansion of the lattice when bromine is progressively replaced by iodine, evidenced by the overall shifting of all characteristic peaks to higher angles.

Thermal expansion trends are very similar for all mixed compositions (see **Figure 6**) in line with the behavior measured experimentally.[34,35,58,62] The predicted values for the thermal expansion coefficient (as derived from the slopes of the volume vs. temperature) are $12.6 \times 10^{-5}$ K$^{-1}$ and $12.7 \times 10^{-5}$ K$^{-1}$, which compare well with the experimentally reported data of $12.1 \times 10^{-5}$ K$^{-1}$ and $11.8 \times 10^{-5}$ K$^{-1}$ for CsPbBr$_3$ and CsPbI$_3$, respectively.

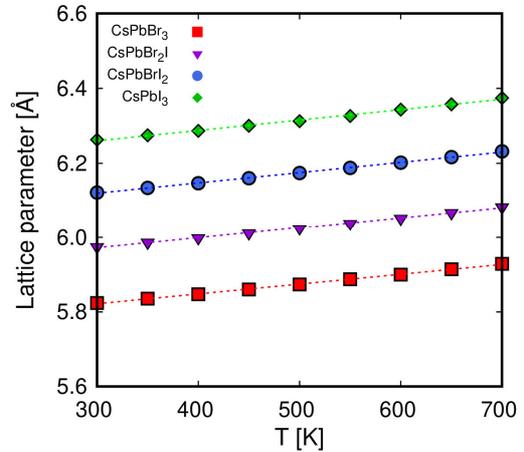

Figure 6. Lattice parameter as a function of temperature for various compositions.

**D: Dynamic properties of mixed perovskites: heat transport and ion migration**

CMD simulation with a suitable force field allows to run dynamics for longer times and larger system sizes than quantum calculations. This way dynamic properties such as thermal conductivities and ionic diffusion coefficients can be obtained for a variety of compositions and structures. As regards the former, in **Figure 7** the ion-mediated heat transfer is computed using the genetically-modified classical force field as a function of time. From the slopes of the curves in the linear regime thermal conductivities for the samples studied can be readily determined. Values of (0.49 ± 0.01) and (0.59 ± 0.04) W m$^{-1}$ K$^{-1}$ were obtained for $CsPbBr_3$ and $CsPbI_3$, respectively. These figures compare remarkably well with measured data (0.45 and 0.42 ± 0.05 W m$^{-1}$ K$^{-1}$),[66] implying that the classical force field correctly reproduces the ultralow thermal conductivity observed experimentally.

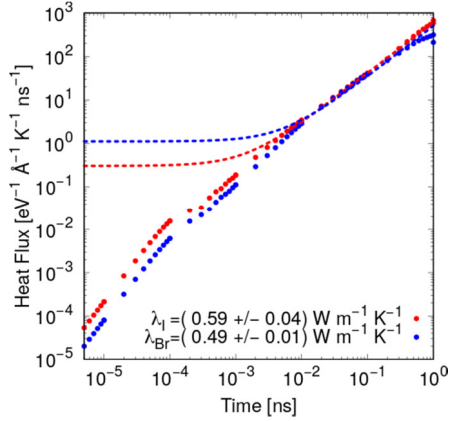

**Figure 7.** Heat flux as a function of time. Solid lines stand for linear fits in the 0.01-0.5 ns interval. The corresponding slopes are the thermal conductivity coefficients (indicated in the graph)

In general it is believed that ion migration plays an important role in explaining the slow response and hysteresis effects that are ubiquitous in the functioning of perovskite solar cells.[67,68] Following previous literature,[11,27,67] and as described in the methodological section, ionic diffusion coefficients are derived from the averaged mean squared displacement of halide ions in structures where a certain concentration of halide vacancies[27,69] are introduced. **Figure 8** shows a 6×6×6 cubic simulation cell where 12 halide ions have been removed at random positions.

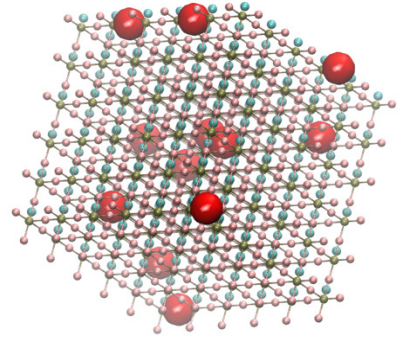

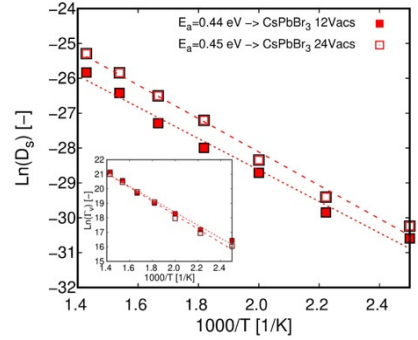

**Figure 8. Top:** 6×6×6 simulation cell with 12 halide vacancies created at random positions. **Bottom:** Ionic diffusion coefficient and ionic jumping rate (inset) as a function of temperature for the $CsPbBr_3$ perovskite derived from CMD simulations at two vacancy concentrations.

CMD simulations of up to 10 ns with the genetically-optimized classical force field were run. The presence of the vacancies allows for ion migration and the generation of a "measurable" mean squared displacement (MSD) according to Eq. (3). Simulation with a perfect crystalline structure (with no vacancies) does not produce MSDs that increase monotonically with time in the 10 ns time window. Hence, swapping of atoms between their respectively equilibrium positions was not observed in this time window in the CMD simulation.

**Figure 9** shows MSD data for the simulations with 12 and 24 vacancies and several temperatures. Two regimes are detected: absence of measurable motion below 0.01-0.1 ns and diffusional transport at longer times ($\alpha$ = 1 in Eq. (4)). The transition between the two regimes strongly depends on temperature: the higher the temperature, the faster the transition from the non-diffusive to the diffusive regime. As we get closer to the ambient temperature of 300 K, diffusion tends to be observed at very long times only, even beyond the time window of 10 ns. As expected, doubling the concentration of vacancies generally accelerates the dynamics, although this effect becomes almost negligible at low temperatures.

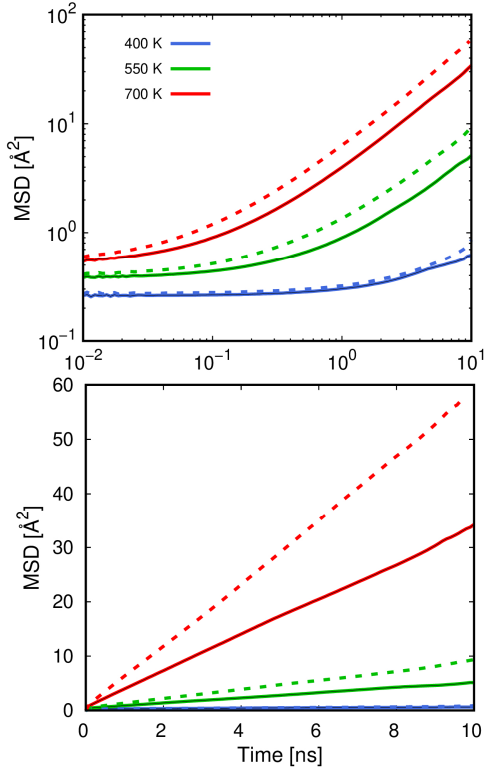

**Figure 9.** Ionic mean squared displacement produced by the CMD simulations with 12 and 24 vacancies per simulation box (solid and dashed lines, respectively) in linear and logarithmic scale.

Eq. (4) applied in the diffusional regime (linear dependence in MSD) provides values for the ionic diffusion coefficient. Results are presented in **Figure 8** and **10**. In line with the trends found in the MSD data, the diffusion coefficient increases with vacancy concentration. In this respect it is important to distinguish between thermodynamic influences on the ionic migration (vacancy concentration) and kinetic factors (energy barriers for ion jumps between available lattice sites). Following Barboni and Souza[27] we have separated both influences via the expression

$$D_i = \frac{a_{\text{geom}} d_i^2 \Gamma_i N_v}{N_i} \quad (5)$$

where $a_{\text{geom}}$ is a geometric factor that depends on the geometry of the crystalline lattice, $d_i$ is the jumping distance for ions, $\Gamma_i$ the jumping rate and $N_v$, $N_i$ are the vacancy and the iodide concentrations, respectively. Eq. **(5)** shows that ionic conduction is faster as more vacancies are available for ionic motion and explains why there is no measurable migration when $N_v = 0$. Using Eq. **(5)** we can renormalize the ionic diffusion coefficients with respect to vacancy concentration and extract pure kinetic jumping rates. Results are collected in the inset of Figure 8. The decoupling of the two effects allows us to conclude that the resulting jumping rates are essentially independent of the vacancy concentration and therefore represent the kinetic barrier for a single vacancy to move through the crystalline lattice. The vacancy jump rate can be related to the enthalpy and entropy change associated to the migration of a single vacancy via,[27]

$$\Gamma_i = \nu_0 \exp\left(\frac{\Delta S_{\text{mig}}}{k_B}\right) \exp\left(\frac{\Delta H_{\text{mig}}}{k_B T}\right) \quad (6)$$

where $\nu_0$ is the phonon attempt-to-jump frequency. Results of the application of Eq. **(6)** along with the activation energies extracted from the diffusion coefficient are collected in **Table 3**. As expected, activation diffusional energies and migration enthalpies are very similar. This indicates that the temperature dependence of the diffusion coefficient yielded by the CMD simulation stems from the kinetic barrier that vacancies have to surmount to migrate through the crystalline lattice. However, it must be noted that in a real situation there is an additional thermal effect produced by the formation of the vacancies itself. This effects appears to be important in methylammonium lead iodide perovskites.[69] The thermal formation of vacancies cannot be considered in the present simulations because their concentration is fixed in the calculation.

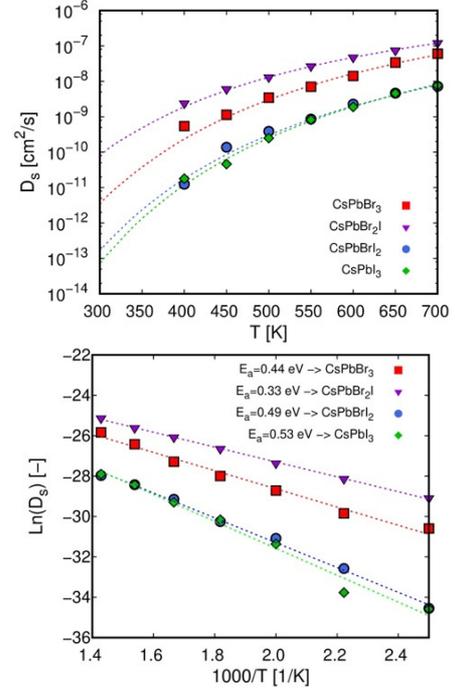

**Figure 10.** Ionic diffusion coefficients from CMD simulation with 12 vacancies for the genetically optimized classical potential and various compositions of the CsPb(Br$_x$I$_{1-x}$)$_3$ perovskite set.

**Table 3.** Thermodynamic parameters from Arrhenius analysis of the temperature dependences of diffusion coefficients and jumping rates.

| Perovskite | $\Delta H_{diff}$ / eV | $\Delta H_{mig}$ / eV | $k_B \Delta S_{mig}$ / eV K$^{-1}$ |
|---|---|---|---|
| CsPbBr$_3$ | 0.44 | 0.43 | -2.38 |
| CsPbBr$_2$I | 0.33 | 0.32 | -3.69 |
| CsPbBrI$_2$ | 0.49 | 0.44 | -5.24 |
| CsPbI$_3$ | 0.53 | 0.55 | -2.07 |

It is also interesting to compare the values obtained in the CMD simulation with previous data from the literature. Mizusaki et al.[70] studied ionic conduction in $CsPbBr_3$ by impedance spectroscopy. They reported activation energy values in the 0.25-0.35 eV range. Activation energies from CMD simulations for $CsPbI_3$ lie very close to values estimated from DFT calculations and impedance measurements for partially substituted $CH_3NH_3PbI_3$ with $Cs^+$.[71] On the other hand, the genetically-optimized classical force field predicts lower activation energies and faster diffusion when iodine is completely replaced by bromine. This could explain why bromine perovskites tend to show reduced hysteresis compared to those with iodine.[72] Interestingly, partial substitution of iodine leads to even faster diffusion and lower activation. The faster migration in the mixed perovskite as predicted by CMD simulation explains why this material tends to phase segregate so easily[73] and why $CsPbBr_2I$ based solar cells exhibit stronger hysteresis.[74] Similar non-linear effects have been recently reported by García-Rodríguez et al.[75] for partially substituted $CH_3NH_3PbI_3$ as a function of bromine content.

Given the lack of literature data, the diffusion coefficient for $CsPbI_3$ at ambient temperature can only be compared with the information for the $CH_3NH_3PbI_3$ perovskite. $10^{-13}$ $cm^2/s$ lies one order of magnitude above the DFT-calculated value of Eames at al.[11] and two orders of magnitude above the figure needed by Richardson et al.[67] to reproduce with their drift-diffusion model the hysteresis of $CH_3NH_3PbI_3$ based solar cells. However, direct comparison with experiment is very difficult because, as shown in the present calculations, the diffusion coefficients strongly depend on the concentration of vacancies, and this is in turn determined by temperature[69], composition and preparation conditions.

## Conclusions

A genetic algorithm has been used to obtain a classical force field valid for all-inorganic metal-halide perovskites of varying composition, i.e., $CsPb(Br_xI_{1-x})_3$, with $x$ ranging between 0 and 1. DFT calculations were used as reference data to find the parameters of the force field that best reproduce the energies of a set of geometric distortions of the lattice. The newly derived force field can reproduce, in good agreement with experiments and previous theoretical findings, the crystalline structure (lattice parameters, thermal expansion coefficients, XRD spectra) and the dynamical properties (thermal conductivities, ionic diffusion coefficients) for all compositions, and at various temperatures, with a single set of parameters. In particular, for vacancy concentrations of around $2.6 \times 10^{20}$ $cm^{-3}$, ionic diffusion coefficients at room temperature of $10^{-11}$ and $10^{-13}$ $cm^2/s$ are estimated for $CsPbBr_3$ and $CsPbI_3$, respectively. In addition, we observe a significant lowering of the activation energy for vacancy migration (and faster diffusion) for the mixed perovskites with respect to the pure ones. The proposed procedure and the effects observed pave the way to obtain transferable classical force fields for other types of perovskites and it further enables efficient screening of optimal compositions and structures with desired dynamical properties for optimized photovoltaic operation. Further work to extend the genetic algorithm to obtain universal, transferable and efficient classical force fields for these and other types of perovskites are currently underway.

## Conflicts of interest

There are no conflicts to declare

## Acknowledgements

We thank Ministerio de Ciencia e Innovación of Spain, Agencia Estatal de Investigación (AEI) and EU (FEDER) under grants MAT2016-79866-R and PCI2019-111839-2 (SOLAR COFUND 2, project SCALEUP) for financial support. We thank Junta de Andalucía for support under grant SOLARFORCE (UPO-1259175). S.T. acknowledges funding by the Computational Sciences for Energy Research (CSER) tenure track program of Shell and NWO (Project number 15CST04-2), the Netherlands. S.R.G.B. thanks Spanish Ministerio de Educación, Cultura y Deporte for his predoctoral and postdoctoral fellowship (BES-2014-067825 from CTQ2013-48396-P). We also thank C3UPO for the HPC support.